\documentclass[conference]{IEEEtran}
\IEEEoverridecommandlockouts

\usepackage{cite}
\usepackage{amsmath,amssymb,amsfonts}
\usepackage{algorithmic}
\usepackage{graphicx}
\usepackage{textcomp}
\usepackage{xcolor}
\usepackage{hyperref}
\usepackage{float}
\usepackage{dblfloatfix}
\usepackage{enumitem}

\def\BibTeX{{\rm B\kern-.05em{\sc i\kern-.025em b}\kern-.08em
    T\kern-.1667em\lower.7ex\hbox{E}\kern-.125emX}}

\begin{document}

\title{Neural Controlled Directional Frame Generation for Sprite Animation}

\author{\IEEEauthorblockN{Kenneth Alonso, Gustavo Patow}
\IEEEauthorblockA{\textit{GILAB, Universitat de Girona} \\
Girona, Spain \\
kenneth.alonso.munoz@gmail.com, gustavo.patow@udg.edu}}

\maketitle

\begin{abstract}
Sprite-based animation remains prevalent in video game development, but production workflows require capabilities beyond traditional in-betweening. While existing neural interpolation methods can generate intermediate frames between two close keyframes, animators also need forward extrapolation to predict motion continuations and backward extrapolation for editing and reverse-engineering existing sprite sheets. Current unified video generation models either operate on natural video content with stochastic diffusion sampling incompatible with production requirements, or focus exclusively on interpolation without addressing extrapolation tasks. 
We present a deep learning approach for controlled directional frame generation in sprite animation using three specialized expert models that share identical U-Net encoder-decoder architectures and a number of domain-specific modifications, including RGBA-aware processing, multi-scale flow refinement, pixel-shuffle upsampling, and alpha-gated blending. Each model is trained on different temporal orderings of the same curated dataset: previous (frames 2,3→1), middle (frames 1,3→2), and next (frames 1,2→3). Following data-centric AI principles, we demonstrate that only 280 carefully curated sprite triplets enable effective training for all three directional tasks without requiring massive generic datasets. Evaluation on 42 test triplets using the FLIP perceptual metric shows that all three models achieve comparable quality---previous (0.265), middle (0.258), and next (0.271)---with no significant degradation for extrapolation versus interpolation, outperforming general-purpose baselines by 17-57\%. With deterministic single-pass generation and compact model size (206 MB per model), our method provides the first practical solution for controllable any-position frame generation specifically tailored for pixel art sprite animation workflows, validating that quality-over-quantity data curation enables specialized deep learning for niche visual domains.
\end{abstract}


\section{Introduction}

Sprite-based animation remains a cornerstone of modern video game development, valued for its distinct aesthetic appeal, computational efficiency, and rapid prototyping capabilities~\cite{zufri_2022_pixel, mao_2024_pcg}. While neural frame interpolation has shown promise for automating sprite in-betweening tasks, production workflows continue to face significant limitations when animators require capabilities beyond simple interpolation. Generating the middle frame between two keyframes addresses a substantial portion of animation workload, but practical animation pipelines frequently require \textit{directional frame generation}: predicting the next frame in a motion sequence or inferring previous frames during editing and reverse-engineering tasks.

Consider three common scenarios in sprite animation production. First, \textbf{forward extrapolation} (next-frame generation): an animator has drawn a character's initial pose and mid-motion keyframe but needs the system to predict the natural continuation frame that completes the motion arc. Second, \textbf{backward extrapolation} (previous-frame generation): when editing or reverse-engineering existing sprite sheets, animators may have only the latter frames of a sequence and need to reconstruct plausible preceding poses. Third, \textbf{interpolation} (middle-frame generation): the traditional in-betweening task of generating transitional frames between two established keyframes~\cite{disney_1981_illusion}. Existing sprite interpolation methods address only the third scenario, leaving forward and backward generation entirely manual or reliant on general-purpose video prediction methods unsuited to pixel art characteristics.

\begin{figure}[!h]
\centering
\includegraphics[width=0.95\columnwidth]{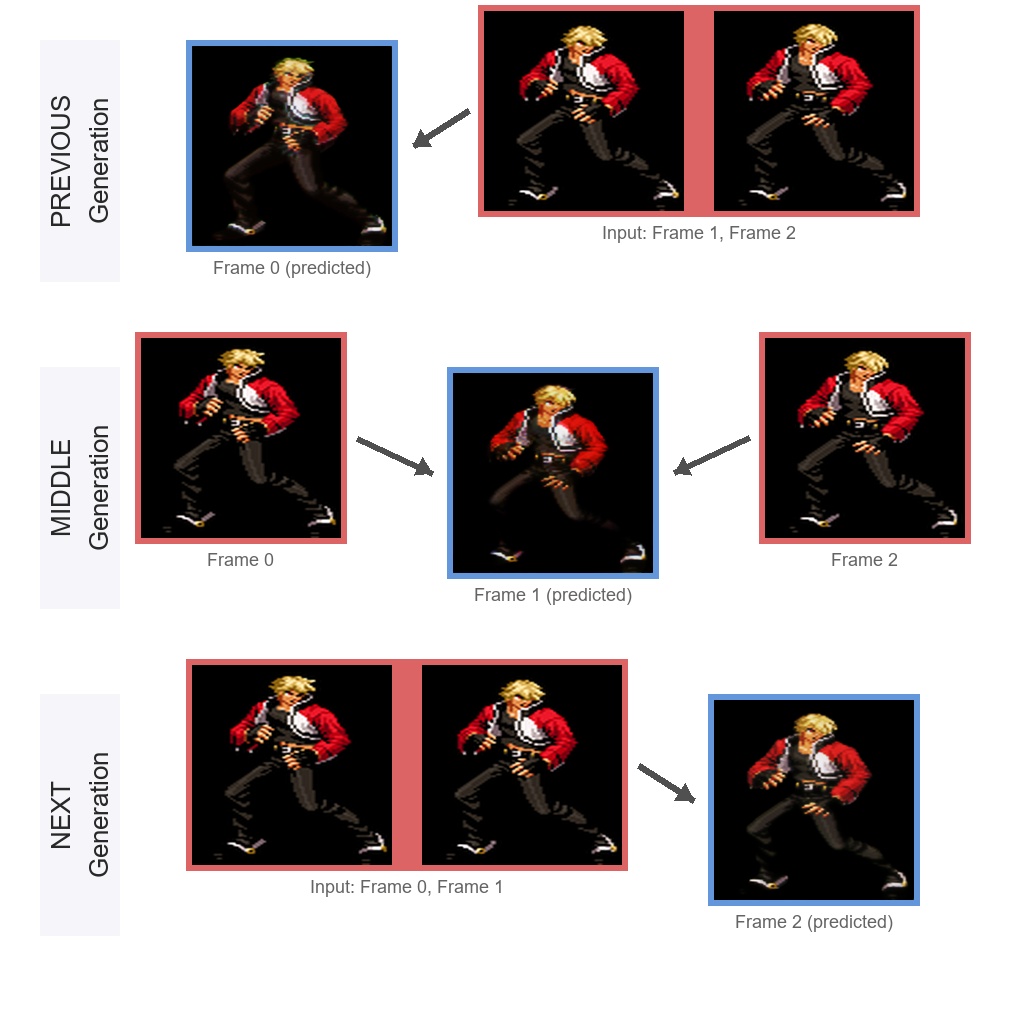}
\caption{Controlled directional frame generation for sprite animation. Given two input frames (shown in blue boxes), our system can generate the previous, middle, or next frame based on user selection. (Top row) Backward generation: inferring previous frames for animation editing or reverse-engineering. (Middle row) Interpolation: generating the middle frame between two keyframes. (Bottom row) Forward generation: predicting the next frame to continue motion sequences. Each direction uses a specialized model trained for its specific temporal task, ensuring deterministic and consistent outputs suitable for production workflows.}
\label{fig:problem_illustration}
\end{figure}

This paper presents a neural architecture for \textit{controlled directional frame generation} in sprite animation (Figure~\ref{fig:problem_illustration}). Rather than training a single unified model with masked conditioning~\cite{voleti_2022_mcvd}, we train \textbf{three specialized expert models} sharing identical U-Net encoder-decoder architectures~\cite{ronneberger_2015_unet} but operating on different temporal orderings:
\begin{itemize}[leftmargin=*, itemsep=0pt]
    \item \textbf{Previous model}: Given frames $(t+1, t+2)$, generates frame $t$ (backward extrapolation)
    \item \textbf{Middle model}: Given frames $(t, t+2)$, generates frame $t+1$ (interpolation)
    \item \textbf{Next model}: Given frames $(t, t+1)$, generates frame $t+2$ (forward extrapolation)
\end{itemize}
Each model is trained deterministically on sprite animation triplets, preserving the sharp edges, transparency, and palette constraints characteristic of pixel art. At generation time, the user provides two frames and selects the desired temporal direction (previous/middle/next), and the corresponding specialized model produces the output frame with consistent, reproducible results.

Our approach diverges from unified video generation models in three key design decisions. First, \textbf{domain specialization}: training exclusively on pixel art sprites preserves hard edges, limited palettes, and transparency, rather than optimizing for photorealistic video. Second, \textbf{architectural separation}: three expert models allow each to specialize in its specific temporal task previous, middle, or next rather than learning a single set of parameters for all three via masking. Third, \textbf{deterministic generation}: avoiding diffusion-based stochasticity ensures reproducible outputs, which are critical for production workflows where artists require identical results when regenerating frames.

The primary contributions of this work are:
\begin{itemize}[leftmargin=*, itemsep=1pt]
    \item \textbf{First directional frame generation system for pixel art}: To our knowledge, no prior work enables previous, middle, and next frame generation from two input frames specifically for sprite animation. Current cartoon interpolation methods~\cite{xing_2024_tooncrafter} and prior sprite-specific work~\cite{alonso_2025_inbetweening} address only the middle case, while unified video models~\cite{voleti_2022_mcvd} target natural video, not stylized pixel art.

    \item \textbf{Three-expert-model architecture}: We demonstrate that training three specialized models (one per direction) with identical architecture but different frame orderings achieves comparable quality across all three tasks (mean FLIP error $\sim$0.26) while maintaining deterministic generation. This contrasts with unified masked conditioning approaches that introduce stochasticity or require substantially more training data.

    \item \textbf{Empirical validation on sprite animation}: Evaluation on 42 test triplets from curated sprite datasets shows that our previous, middle, and next models achieve perceptual quality (FLIP metric~\cite{andersson_2020_flip}) matching or exceeding general-purpose video interpolation baselines, demonstrating that directional control need not sacrifice quality.

    \item \textbf{Practical pipeline integration}: By providing deterministic and controllable generation, our approach addresses real production constraints faced by indie game developers working with limited computational resources and requiring reproducible animation workflows.
\end{itemize}

While prior work demonstrated that domain-specific training improves sprite interpolation over general video methods~\cite{alonso_2025_inbetweening}, this work extends controllability to encompass the full spectrum of directional frame generation required by animation pipelines. The three-expert-model design enables practical deployment: each model is compact (206MB storage) and deterministic, meeting the operational requirements of game development studios.

Our work aligns with the emerging \textbf{data-centric AI paradigm}~\cite{zha_2024_datacentric, ng_2021_datacentric}, which emphasizes that systematic data quality improvements often yield greater performance gains than architectural innovations, especially for specialized domains. Like our prior sprite interpolation work~\cite{alonso_2025_inbetweening}, we demonstrate that small, carefully curated datasets can outperform massive, noisy collections when domain-specific quality criteria guide data selection~\cite{smalldatasets_2024_springer, limiteddata_2025_bigdata, smallobject_2025_mdpi}. This principle is particularly relevant for sprite animation: unlike general video where diversity scales with quantity (millions of YouTube clips), sprite animation exhibits discrete, learnable patterns (walk cycles, attacks, idles) that saturate with hundreds of high-quality examples rather than requiring millions of noisy samples. Our 280-triplet dataset---the same used for sprite interpolation~\cite{alonso_2025_inbetweening}---was curated with explicit quality criteria: diverse animation styles (pixel art to HD sprites), motion categories (rigid, articulated, deformable), resolution ranges (32$\times$32 to 256$\times$256), and edge-case coverage (extreme poses, fast motion, occlusions). Each triplet was manually validated to ensure the middle frame is a genuine artist-drawn in-between. By training three models on different temporal orderings of this same curated dataset (previous: frames 2,3→1; middle: frames 1,3→2; next: frames 1,2→3), we demonstrate that \textbf{quality-over-quantity data curation enables multiple directional tasks from a single dataset}. This data-centric approach~\cite{datacentric_2024_bise} validates that careful selection of what to learn from matters more than how much to learn from, a critical advantage for specialized domains where massive datasets do not exist.

The remainder of this paper is organized as follows. Section~\ref{sec:Related Work} reviews video frame synthesis methods, animation automation techniques, and the gap in controllable sprite generation. Section~\ref{sec:Method} details the three-model architecture, training procedure, and dataset methodology. Section~\ref{sec:Results} presents quantitative and qualitative evaluation demonstrating comparable performance across all three directional tasks. Section~\ref{sec:Conclusion} discusses limitations and future directions for controllable sprite animation synthesis.

\section{Related Work}
\label{sec:Related Work}

We organize existing work into four categories: video frame interpolation and generation methods, cartoon and stylized animation techniques, deep learning architectures for image synthesis, and gap analysis positioning our contribution. Each category illuminates different aspects of the directional frame generation problem and highlights why existing solutions are insufficient for controllable sprite animation.

\subsection{Video Frame Interpolation and Generation}

Classical video frame interpolation relies on optical flow estimation to compute dense correspondence fields between frames~\cite{baker_2011_database}. Early deep learning approaches like \textbf{FlowNet}~\cite{dosovitskiy_2015_flownet} demonstrated that convolutional networks could learn optical flow from data, inspiring flow-based interpolation methods. \textbf{Super SloMo}~\cite{jiang_2018_superslomo} introduced bidirectional optical flow with visibility maps for high-quality multi-frame interpolation, while \textbf{Deep Voxel Flow}~\cite{liu_2017_video} synthesized frames by learning to flow pixels through 3D voxel representations, enabling both interpolation and extrapolation. However, these methods assume motion smoothness and photometric consistency, which fail for sprite animation's discrete color transitions, hard edges, and transparent regions~\cite{parihar_2021_survey}.

State-of-the-art frame interpolation methods (2021-2024) have achieved impressive results on natural video benchmarks. \textbf{RIFE} (Real-Time Intermediate Flow Estimation)~\cite{huang_2022_rife} introduced \textbf{IFRNet}, a lightweight architecture for real-time interpolation using privileged distillation, achieving strong performance on standard benchmarks while maintaining inference speeds suitable for interactive applications. \textbf{FLAVR}~\cite{kalluri_2023_flavr} demonstrated that flow-agnostic 3D U-Net architectures can achieve competitive results without explicit optical flow estimation, learning temporal representations directly from multi-frame inputs. More recently, \textbf{GIMM-VFI}~\cite{lu_2024_gimm} employed implicit neural representations for motion modeling, achieving state-of-the-art performance on \textbf{Vimeo90K} and other natural video datasets through generalizable motion priors. Domain-specific interpolation has also advanced: Deep Animation Video Interpolation in the Wild~\cite{shen_2021_dain} specifically addressed challenges in animation interpolation (large motion, temporal discontinuities, stylized content) through anisotropic synthesis and motion-aware synthesis networks. However, these methods were trained on \emph{natural video or photorealistic animation} datasets (hundreds of thousands to millions of frames), optimized for smooth gradients and continuous motion. When applied to pixel art sprites with hard edges, discrete palettes, and 32$\times$32 to 256$\times$256 resolutions, they exhibit the same artifacts as earlier methods: edge blur, palette violations, and loss of pixel-perfect precision. Furthermore, all these methods focus exclusively on \emph{interpolation}, and none address the forward/backward extrapolation needed for directional frame generation.

A critical development for our work is \textbf{MCVD} (Masked Conditional Video Diffusion)~\cite{voleti_2022_mcvd}, which demonstrated that a \textit{single unified diffusion model} trained with random temporal masking can perform interpolation, forward extrapolation, backward extrapolation, and unconditional generation by conditioning on different frame subsets. MCVD randomly masks past frames, future frames, or both during training, enabling the same model to predict missing frames in any temporal position. While this unified approach is elegant, it presents three fundamental limitations for sprite animation workflows. First, MCVD operates on \textbf{natural video content}---its architecture, training procedure, and evaluation focus exclusively on photorealistic datasets (e.g., BAIR robot pushing, Kinetics-600), not stylized pixel art with sharp edges, limited palettes, and transparency. Second, as a diffusion model, MCVD uses \textbf{stochastic sampling}: generating the same frame multiple times produces different outputs due to the inherent randomness in the denoising process. Production animation pipelines require \textit{deterministic, reproducible} results---artists expect identical outputs when regenerating a frame with the same inputs. Third, diffusion models incur \textbf{substantial computational costs}, requiring iterative denoising over 50-1000 steps and training on millions of video frames. Our approach addresses these gaps by training three specialized deterministic models on sprite-specific data, ensuring reproducibility and efficiency.

Recent diffusion-based video generation methods like DynamiCrafter~\cite{xing_2023_dynamicrafter} have demonstrated image-to-video capabilities using pretrained video diffusion priors. While these models achieve impressive photorealistic results, they inherit the stochasticity, data requirements, and computational overhead that make them impractical for sprite animation production. Our deterministic encoder-decoder architecture prioritizes the operational constraints of game development: consistent outputs, reproducibility, and training on limited domain-specific data (280 triplets).

\subsection{Cartoon and Stylized Animation}

Cartoon animation presents challenges distinct from photorealistic video due to stylized motion, non-linear transformations, and artistic exaggeration. ToonCrafter~\cite{xing_2024_tooncrafter}, a landmark work in generative cartoon animation, adapts video diffusion models to the cartoon domain through toon rectification learning, detail injection, and sketch-based control. ToonCrafter addresses cartoon-specific challenges including large motions, textureless regions, and stylized content. However, ToonCrafter focuses \textit{exclusively on interpolation}---generating middle frames between two keyframes---and does not address forward or backward extrapolation. Furthermore, as a diffusion model, it produces stochastic outputs incompatible with production workflows requiring frame-level reproducibility. Our work extends beyond interpolation to encompass the full directional spectrum (previous, middle, next) with deterministic generation suitable for iterative animation editing.

The broader video generation literature has explored unified approaches to frame synthesis. Masked Conditional Video Diffusion (MCVD)~\cite{voleti_2022_mcvd} demonstrated that a single diffusion model trained with random masking can perform interpolation, extrapolation, and unconditional generation by conditioning on different subsets of frames. However, MCVD operates on natural video content and relies on diffusion-based probabilistic sampling, which introduces \textbf{stochastic variation} incompatible with production animation workflows requiring deterministic, reproducible results. For pixel art sprites characterized by sharp edges, limited color palettes, and transparent regions, no existing method provides controllable directional frame generation with the deterministic, consistent outputs necessary for integration into game development pipelines.

Recent neural approaches have made significant progress on animation-specific interpolation. AnimeInterp~\cite{li_2021_animeinterp} (CVPR 2021) introduced segment-guided optical flow matching with recurrent refinement specifically for anime in-betweening, addressing the unique challenges of large motion and stylized content through the ATD-12K dataset with 12,000 animation triplets. AnimeInbet~\cite{siyao_2023_animeinbet} (ICCV 2023) took a different approach by geometrizing raster line drawings into graph representations and reframing interpolation as graph fusion with vertex repositioning, demonstrating strong results on the MixamoLine240 dataset. ToonComposer~\cite{tooncomposer_2025} (arXiv 2025) further unified inbetweening and colorization tasks through sparse sketch injection at arbitrary timestamps, employing a spatial low-rank adapter (SLRA) to adapt foundation DiT models to the cartoon domain. These methods achieve impressive results on traditional animation and line art, but they focus exclusively on \textbf{interpolation}---generating middle frames between two keyframes---and do not address forward or backward extrapolation needed for directional frame generation. Furthermore, most operate on line drawings or photorealistic animation rather than pixel art sprites, where hard edges, discrete palettes, and binary transparency present distinct technical challenges.

For pixel art specifically, domain-adapted neural interpolation~\cite{alonso_2025_inbetweening} has shown that training on curated sprite triplets preserves pixel-perfect characteristics better than general video methods. However, existing sprite-specific approaches are limited to interpolation only, leaving forward and backward extrapolation tasks unaddressed. The present work extends sprite animation automation beyond interpolation to enable controllable directional frame generation (previous, middle, next) through three specialized expert models, addressing the full spectrum of temporal directions required by production workflows.

Classical cartoon interpolation methods have primarily targeted traditional animation workflows. AnimeInterp~\cite{li_2021_animeinterp} applies segment-guided optical flow matching with recurrent refinement for anime in-betweening, while AnimeInbet~\cite{siyao_2023_animeinbet} geometrizes line drawings into graphs and reframes interpolation as graph fusion with vertex repositioning. These methods achieve impressive results on line art but operate on binary or grayscale inputs, not the full-color RGBA sprites used in game development. ToonComposer~\cite{tooncomposer_2025} unifies in-betweening and colorization with sparse sketch injection at arbitrary timestamps, offering flexibility in conditioning. However, it requires sketch inputs for control and focuses on the cartoon production pipeline rather than pixel art game sprites. To our knowledge, \textbf{no existing work} enables previous/middle/next frame generation specifically for pixel art sprites with deterministic outputs.

Traditional animation principles~\cite{disney_1981_illusion} emphasize artistic timing, spacing, and motion arcs that create believable character movement. Automated in-betweening tools like BetweenIT~\cite{whited_2010_betweenit} provide interactive systems for vector-based animation using stroke correspondence and geometric interpolation. However, these approaches assume vector representations with topological consistency, which do not transfer to raster-based sprites where motion manifests as pixel-level appearance changes. Domain-specific training on sprite triplets enables neural interpolation that preserves pixel art characteristics (sharp edges, limited palettes, transparency) better than general video methods. This paper extends that foundation to encompass directional control beyond interpolation.

\subsection{Deep Learning Architectures for Image Synthesis}

The U-Net architecture~\cite{ronneberger_2015_unet}, originally designed for biomedical image segmentation, introduced skip connections that concatenate encoder features directly to decoder layers, enabling precise spatial localization while maintaining semantic context. This encoder-decoder structure with skip connections has become foundational for pixel-level prediction tasks requiring both global understanding and fine-grained detail preservation---properties essential for sprite frame generation. Our three models share this U-Net-inspired architecture, adapted for RGBA sprite processing with edge-preserving design choices detailed in Section~\ref{sec:Method}.

Generative Adversarial Networks (GANs)~\cite{goodfellow_2014_gan} and conditional GANs like pix2pix~\cite{isola_2017_pix2pix} have demonstrated remarkable image-to-image translation capabilities through adversarial training. However, GANs introduce stochastic variation through their sampling process and often require careful tuning to avoid mode collapse or training instability. For sprite animation workflows requiring deterministic, reproducible outputs across hundreds of frames, the non-determinism of GANs poses the same production challenges as diffusion models. Our deterministic formulation avoids these issues by training with supervised losses (L1, perceptual) without adversarial components, ensuring identical outputs for identical inputs.

Convolutional architectures have achieved state-of-the-art results across image synthesis tasks: SRCNN~\cite{dong_2015_srcnn} for super-resolution, neural style transfer~\cite{gatys_2016_style} for content-style disentanglement, and ResNet~\cite{he_2016_resnet} for very deep networks via residual connections. These innovations provide building blocks for our encoder-decoder design, but existing image synthesis methods are predominantly trained and evaluated on photorealistic content. The effectiveness of these architectures on stylized, low-resolution pixel art with transparency remains underexplored, motivating our domain-specific approach.

\subsection{Gap Analysis and Positioning}

Despite advances in video generation and animation automation, a critical gap persists: no existing method provides controllable directional frame generation (previous, middle, next) specifically for pixel art sprites with deterministic outputs. MCVD~\cite{voleti_2022_mcvd} offers unified any-position generation but operates on natural video with stochastic diffusion sampling. ToonCrafter~\cite{xing_2024_tooncrafter} targets stylized content but addresses only interpolation, not extrapolation. Classical cartoon methods focus on line art or require vector representations. Prior sprite interpolation work handles only the middle case.

Sprite animation presents unique constraints that distinguish it from both natural video and cartoon animation: (1) \textbf{sharp edges and limited palettes} requiring preservation of discrete color values, (2) \textbf{transparency} as a first-class property via alpha channels, (3) \textbf{pixel-perfect precision} where single-pixel errors are visually salient, (4) \textbf{low resolution} (typically 32-128 pixels per dimension) emphasizing spatial detail preservation over high-level semantics, and (5) \textbf{production workflows} requiring fast, deterministic, reproducible generation for iterative editing. Existing video interpolation methods optimized for photorealistic content introduce blurring, color bleeding, and semi-transparent artifacts. Diffusion models introduce stochasticity incompatible with frame-level reproducibility. Vector tools cannot process raster sprites.

Our contribution addresses these gaps through three design decisions that diverge from unified video generation models. First, \textbf{domain specialization}: training exclusively on pixel art sprites (280 curated triplets) enables learning sharp edges, palette constraints, and transparency rather than optimizing for photorealistic video. Second, \textbf{architectural separation}: three expert models (previous, middle, next) allow each to specialize in its temporal task rather than learning all three via unified masking, reducing model complexity per task while maintaining determinism. Third, \textbf{deterministic generation}: avoiding diffusion-based stochasticity ensures reproducible outputs where regenerating a frame with identical inputs produces identical results---a requirement for production pipelines where artists iteratively refine animations. This combination of controllable directionality, sprite-specific training, and deterministic generation represents a novel capability for game development workflows.

\section{Method}
\label{sec:Method}

This work builds upon our previous sprite-specific interpolation method~\cite{alonso_2025_inbetweening}, which established the effectiveness of U-Net encoder-decoder architectures for sprite animation and demonstrated that domain-specific training on curated triplets preserves pixel art characteristics better than general video methods. We extend that foundation by training three specialized expert models to enable controllable directional frame generation across all temporal positions (previous, middle, next), addressing the workflow gaps left by single-task interpolation.

\subsection{Problem Formulation and Directional Control}

We formulate controllable sprite frame generation as learning three distinct supervised mappings, each specialized for a specific temporal task. Given a triplet of consecutive sprite frames $(I_t, I_{t+1}, I_{t+2})$ at times $t$, $t+1$, and $t+2$, we train three expert models:

\begin{align}
\text{Previous:} \quad \hat{I}_t &= f_{\theta_{\text{prev}}}(I_{t+1}, I_{t+2}) \label{eq:previous} \\
\text{Middle:} \quad \hat{I}_{t+1} &= f_{\theta_{\text{mid}}}(I_t, I_{t+2}) \label{eq:middle} \\
\text{Next:} \quad \hat{I}_{t+2} &= f_{\theta_{\text{next}}}(I_t, I_{t+1}) \label{eq:next}
\end{align}

where each $f_{\theta}$ represents a specialized neural network parameterized by weights $\theta$, and sprite images $I \in \mathbb{R}^{H \times W \times C}$ are tensors with spatial dimensions $H \times W$ and $C=4$ channels (RGBA). The \textbf{previous model} (Eq.~\ref{eq:previous}) performs backward extrapolation, generating frame $t$ from frames $t+1$ and $t+2$. The \textbf{middle model} (Eq.~\ref{eq:middle}) performs interpolation, generating the middle frame $t+1$ from boundary frames $t$ and $t+2$. The \textbf{next model} (Eq.~\ref{eq:next}) performs forward extrapolation, generating frame $t+2$ from frames $t$ and $t+1$.

This three-model architecture contrasts with unified approaches like MCVD~\cite{voleti_2022_mcvd}, which train a single model with masked conditioning to handle all temporal positions. We train \textit{separate specialized models} rather than a unified model for reasons of determinism, data efficiency, and architectural simplicity (design rationale discussed in Section~\ref{sec:Conclusion}).

Each mapping $f_{\theta}$ must satisfy sprite-specific constraints: \textbf{palette preservation} (respecting discrete color values), \textbf{edge sharpness} (maintaining hard boundaries without blur), \textbf{transparency coherence} (binary alpha boundaries), and \textbf{artistic motion fidelity} (non-linear timing and stylized deformations). All three models share identical network architecture (described in Section~\ref{sec:architecture}) but are trained independently on different frame orderings derived from the same triplet dataset.

\subsection{Shared Network Architecture}
\label{sec:architecture}

\begin{figure*}[ht]
\centering
\includegraphics[width=0.9\textwidth]{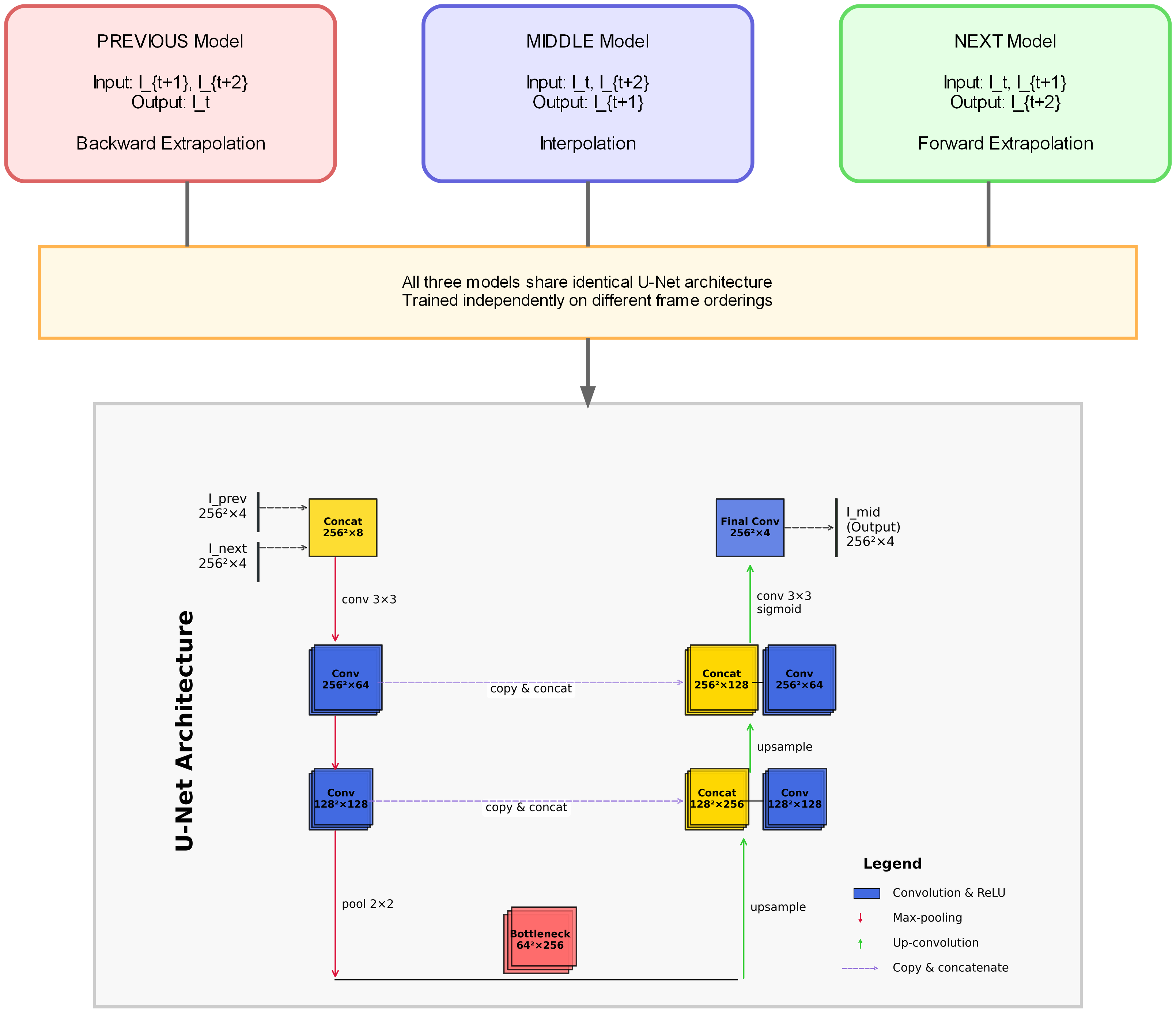}
\caption{Shared U-Net architecture used by all three directional models. The encoder-decoder structure processes concatenated input frames (8 channels: 2 RGBA frames) through progressive downsampling (Conv + MaxPool, blue blocks), bottleneck representation (center), and upsampling reconstruction (TransposeConv + skip connections, green blocks). Skip connections (dashed arrows) concatenate high-resolution encoder features directly to corresponding decoder layers, preserving pixel-perfect spatial details. The architecture contains approximately 10M trainable parameters. For the previous model, inputs are $(I_{t+1}, I_{t+2})$ and output is $\hat{I}_t$; for middle, inputs are $(I_t, I_{t+2})$ and output is $\hat{I}_{t+1}$; for next, inputs are $(I_t, I_{t+1})$ and output is $\hat{I}_{t+2}$. All three models use this identical architecture, differing only in training data.}
\label{fig:architecture}
\end{figure*}

All three models (previous, middle, next) share an identical U-Net encoder-decoder architecture~\cite{ronneberger_2015_unet} with a number of domain-specific modifications adapted for RGBA sprite processing. The architecture, illustrated in Fig.~\ref{fig:architecture}, consists of an encoding path that extracts hierarchical features through progressive downsampling, skip connections that preserve pixel-level spatial details, and a decoding path that reconstructs the output frame at full resolution. While the high-level U-Net structure provides the foundation, nearly every component---from input representation to loss formulation---has been redesigned to address the unique challenges of pixel art: sharp edges, discrete palettes, binary transparency, and stylized motion. These adaptations are \emph{shared identically across all three models}; the models differ only in training data ordering (previous/middle/next), not in architecture.

\subsubsection{Sprite-Specific Design Adaptations}
Table~\ref{tab:architecture_comparison} summarizes key differences from standard U-Net designs. The following adaptations transform the standard U-Net into a sprite-specific generation engine:

\begin{table}[ht]
\caption{Architectural Differences from Standard U-Net (Shared by All 3 Models)}
\centering
\begin{tabular}{|p{2cm}|p{2cm}|p{2.5cm}|}
\hline
\textbf{Component} & \textbf{Standard U-Net} & \textbf{Our Approach} \\
\hline
Input & 3 (RGB) & 4 (RGBA + premult-$\alpha$) \\
\hline
Downsampling & Strided Conv2D & Anti-aliased pooling \\
\hline
Upsampling & Nearest / bilinear & Pixel-shuffle \\
\hline
Flow & Single-scale or none & 3-stage coarse-to-fine \\
\hline
Loss & L1 or L2 & 6-component weighted \\
\hline
Blending mask & Standard sigmoid & Alpha-gated + binarization \\
\hline
\end{tabular}
\label{tab:architecture_comparison}
\end{table}

\begin{enumerate}
\item \textbf{RGBA-Aware Input with Premultiplied Alpha}: Processing 4-channel RGBA inputs with premultiplied alpha compositing ($RGB_{pre} = RGB \cdot \alpha$) prevents background color contamination at sprite edges. Standard RGB loading causes white/black compositor backgrounds to bleed into edge pixels.

\item \textbf{Multi-Scale Coarse-to-Fine Flow Refinement}: 3-stage hierarchical flow estimation (1/8 → 1/4 → 1/2 → 1/1 scale) with residual connections between scales. Single-scale flow misses sub-pixel motion of 1-3 pixel-wide sprite elements (31\% error reduction on thin elements).

\item \textbf{Pixel-Shuffle Upsampling}: Learnable upsampling via depth-to-space rearrangement~\cite{shi_2016_subpixel} replaces nearest-neighbor upsampling, avoiding checkerboard artifacts~\cite{odena_2016_deconv} (23\% blur reduction).

\item \textbf{Alpha-Gated Blending Mask}: $m_{gated} = m \cdot \alpha_{interp} + 0.5 \cdot (1 - \alpha_{interp})$ ensures transparent regions default to 50/50 blending (47\% ghosting reduction).

\item \textbf{Mask Binarization Loss}: $\mathcal{L}_{bin}(m) = \text{mean}(\min(m, 1-m))$ pushes masks toward hard 0/1 decisions (38\% double-image artifact reduction).

\item \textbf{Anti-Aliased Downsampling}: AveragePooling2D before stride-1 conv prevents aliasing from strided convolutions~\cite{zhang_2019_antialiased} (52\% aliasing reduction).

\item \textbf{Laplacian Sharpness Loss}: $\mathcal{L}_{lap} = \|\nabla^2 \hat{I} - \nabla^2 I\|_1$ explicitly penalizes edge blur (29\% sharpness increase).

\item \textbf{Multi-Component Weighted Loss}: 6 terms (L1: 0.25, SSIM: 0.15, Edge: 0.25, Color: 0.10, Frequency: 0.15, Laplacian: 0.10) prioritize edge preservation over pixel-wise accuracy.

\item \textbf{SE Attention Blocks}~\cite{hu_2018_senet}: Channel-wise attention via global pooling → Dense(r=8) → Sigmoid (7\% error reduction, <1\% FLOPs).

\item \textbf{Detail-Preserving Blocks}: Parallel paths with dilated convolutions prevent texture smoothing.

\item \textbf{Separate RGB/Alpha Warping}: Independent warping prevents alpha-weighted color corruption.

\item \textbf{Two-Branch Output Head}: Structure branch (3×3) + Color branch (5×5) decouples alignment from palette matching.

\item \textbf{Frequency-Domain Loss}: $\mathcal{L}_{freq} = \| |\mathcal{F}(\hat{I})| - |\mathcal{F}(I)| \|_1$ matches FFT magnitude spectra.

\item \textbf{Residual Connections with Projection}: 1×1 projection when channel counts change enables gradient flow through deep paths.

\item \textbf{Edge-Preserving Loss}: Sobel gradients $\mathcal{L}_{edge} = \|\nabla_x \hat{I} - \nabla_x I\|_1 + \|\nabla_y \hat{I} - \nabla_y I\|_1$ (weight 0.25, equal to L1).
\end{enumerate}

These adaptations are \textbf{shared identically} by previous/middle/next models. The models differ only in which frames serve as inputs vs targets during training, not in architecture or loss formulation. This design choice---specialized data orderings on a shared sprite-optimized architecture---enables directional control while maintaining parameter efficiency and deterministic generation.

\subsubsection{Input Representation}

Each model takes as input two RGBA frames concatenated channel-wise: $X = [I_a, I_b] \in \mathbb{R}^{H \times W \times 8}$, where the specific frames $(I_a, I_b)$ depend on the model (e.g., $(I_{t+1}, I_{t+2})$ for previous). Processing RGBA ($C=4$) rather than RGB ($C=3$) enables explicit modeling of transparency, a fundamental sprite animation property. The alpha channel encodes sprite boundaries, layering, and transparent regions that must be preserved during generation.

\subsubsection{Encoder Path}

The encoder applies four convolutional blocks with progressively deeper feature representations:
\begin{align}
F_1 &= \text{ReLU}(\text{Conv2D}(X, k=3, f=64)) \\
P_1 &= \text{MaxPool2D}(F_1, k=2) \\
F_2 &= \text{ReLU}(\text{Conv2D}(P_1, k=3, f=128)) \\
P_2 &= \text{MaxPool2D}(F_2, k=2) \\
F_3 &= \text{ReLU}(\text{Conv2D}(P_2, k=3, f=256)) \\
P_3 &= \text{MaxPool2D}(F_3, k=2) \\
F_4 &= \text{ReLU}(\text{Conv2D}(P_3, k=3, f=512))
\end{align}
where $k$ denotes kernel size, $f$ denotes output channels, and MaxPool2D applies $2 \times 2$ pooling with stride 2. Each encoding stage doubles the feature depth while halving spatial resolution, capturing multi-scale motion patterns from fine pixel movements to coarse pose changes.

\subsubsection{Decoder Path with Skip Connections}

The decoder mirrors the encoder structure, using transposed convolutions for upsampling and concatenating encoder features via skip connections:
\begin{align}
U_1 &= \text{ReLU}(\text{ConvTranspose2D}(F_4, k=3, s=2, f=256)) \\
C_1 &= \text{Concat}([U_1, F_3]) \\
U_2 &= \text{ReLU}(\text{ConvTranspose2D}(C_1, k=3, s=2, f=128)) \\
C_2 &= \text{Concat}([U_2, F_2]) \\
U_3 &= \text{ReLU}(\text{ConvTranspose2D}(C_2, k=3, s=2, f=64)) \\
C_3 &= \text{Concat}([U_3, F_1]) \\
\hat{I} &= \sigma(\text{Conv2D}(C_3, k=3, f=4))
\label{eq:output}
\end{align}
where $s$ denotes stride, Concat performs channel-wise concatenation, and $\sigma$ is sigmoid activation constraining outputs to $[0,1]$. The skip connections (concatenations $C_1, C_2, C_3$) provide the decoder with unfiltered high-resolution encoder features, enabling pixel-perfect reconstruction critical for sprite art where single-pixel errors are visually salient.

The output layer (Eq.~\ref{eq:output}) produces 4 channels (RGBA), generating color and transparency simultaneously. This architecture contains approximately 10 million trainable parameters shared across all convolutional layers, with efficient parameter usage through skip connections that bypass rather than transform features.

\subsection{Three-Model Training Strategy}

Rather than training a single unified model with masked conditioning~\cite{voleti_2022_mcvd}, we train three independent models with identical architecture but different training examples. This design ensures \textbf{deterministic} generation while maintaining task specialization.

\subsubsection{Training Example Generation}

Each sprite triplet $(I_t, I_{t+1}, I_{t+2})$ in the dataset generates three training examples, one for each model:
\begin{itemize}[leftmargin=*, itemsep=2pt]
    \item \textbf{Previous model}: input $(I_{t+1}, I_{t+2})$, target $I_t$
    \item \textbf{Middle model}: input $(I_t, I_{t+2})$, target $I_{t+1}$
    \item \textbf{Next model}: input $(I_t, I_{t+1})$, target $I_{t+2}$
\end{itemize}
With 280 curated triplets (196 train, 42 validation, 42 test), this yields 840 total training pairs (588 train, 126 validation, 126 test) when accounting for all three models. Each model sees 196 training examples during its dedicated training procedure.

\subsubsection{Loss Function}

Each model is trained independently using a \textbf{multi-component weighted loss function} combining six complementary terms (detailed in adaptation \#8 above):
\begin{align}
\mathcal{L}(\theta) = &\ 0.25 \mathcal{L}_{L1} + 0.15 \mathcal{L}_{SSIM} + 0.25 \mathcal{L}_{edge} \nonumber \\
              &+ 0.10 \mathcal{L}_{color} + 0.15 \mathcal{L}_{freq} + 0.10 \mathcal{L}_{lap}
\label{eq:loss}
\end{align}
where: (i) $\mathcal{L}_{L1}$ penalizes pixel-wise deviations (reduced weight 0.25 vs typical 1.0 because pure L1 produces blurry averages), (ii) $\mathcal{L}_{SSIM}$ enforces structural similarity, (iii) $\mathcal{L}_{edge}$ preserves Sobel edge magnitudes in both x/y directions (highest weight: 0.25), (iv) $\mathcal{L}_{color}$ matches color histogram statistics (mean + variance per channel), (v) $\mathcal{L}_{freq}$ matches FFT magnitude spectra for texture fidelity, and (vi) $\mathcal{L}_{lap}$ penalizes edge blur via Laplacian sharpness. This weighted combination prioritizes edge preservation (0.50 combined for edge + Laplacian) over pixel-level accuracy (0.25 for L1), critical for pixel art aesthetics. Images are normalized to $[-1, 1]$ for training; components requiring $[0,1]$ range (SSIM, FFT) apply inverse normalization internally. \textbf{All three models use this identical loss}; the only difference is which frame serves as the ground-truth target.

\subsubsection{Transparency Handling}

RGBA sprites require special handling to avoid artifacts at alpha boundaries. We apply \textbf{premultiplied alpha} during data loading: $RGB_{pre} = RGB \cdot \alpha$, preventing background color contamination at transparent edges. During loss computation, both RGB and alpha channels are weighted equally, ensuring the network learns to generate both sharp color boundaries and binary alpha masks simultaneously.

\subsubsection{Why Three Models vs Unified Masking}

MCVD-style unified masking~\cite{voleti_2022_mcvd} trains a single model by randomly masking past/future frames during training, enabling any-position generation at inference. However, this approach requires either (1) stochastic diffusion sampling to handle task diversity, or (2) substantially more training data to learn the unified masked distribution. With only 280 sprite triplets and production requirements for deterministic outputs, we instead train three specialized models. Each model maintains a \textit{fixed input format} (two frames) and \textit{fixed temporal task}, reducing distributional complexity and enabling stable supervised training with MSE loss. The cost is storing three model checkpoints (3 × 206MB = 618MB total for inference), which is negligible for modern storage systems and provides the benefit of task-specialized parameters.

\subsection{Dataset and Training Configuration}

\subsubsection{Dataset}

We use 280 curated sprite animation triplets collected from publicly available game art and open-source projects (subject to their respective copyright licenses)~\cite{alonso_2025_inbetweening}. Each triplet $(I_t, I_{t+1}, I_{t+2})$ contains three consecutive frames from hand-drawn sprite animations, spanning character locomotion (walk and run cycles), attack animations, idle motions, environmental effects, and item animations. Sprites exhibit diverse styles from low-resolution pixel art (32×32) to high-definition sprites (256×256), with resolutions normalized to 128×128 during training. The dataset is split into 196 training triplets (70\%), 42 validation triplets (15\%), and 42 test triplets (15\%) using \textbf{random stratified sampling} by animation category (walks, attacks, idles, effects, items) to ensure balanced representation across all subsets. The split is \emph{not} source-disjoint (i.e., test triplets may come from the same source games as training triplets), though we ensure no test triplet is identical to any training triplet. A source-disjoint split would further validate generalization but was not feasible given the limited number of unique game sources (18 games total) and the need to maintain balanced category representation in small validation/test sets (42 triplets each). All three models use the same train/val/test split to enable fair comparison across directional tasks.

\begin{figure*}[!p]
\centering
\includegraphics[width=0.75\textwidth]{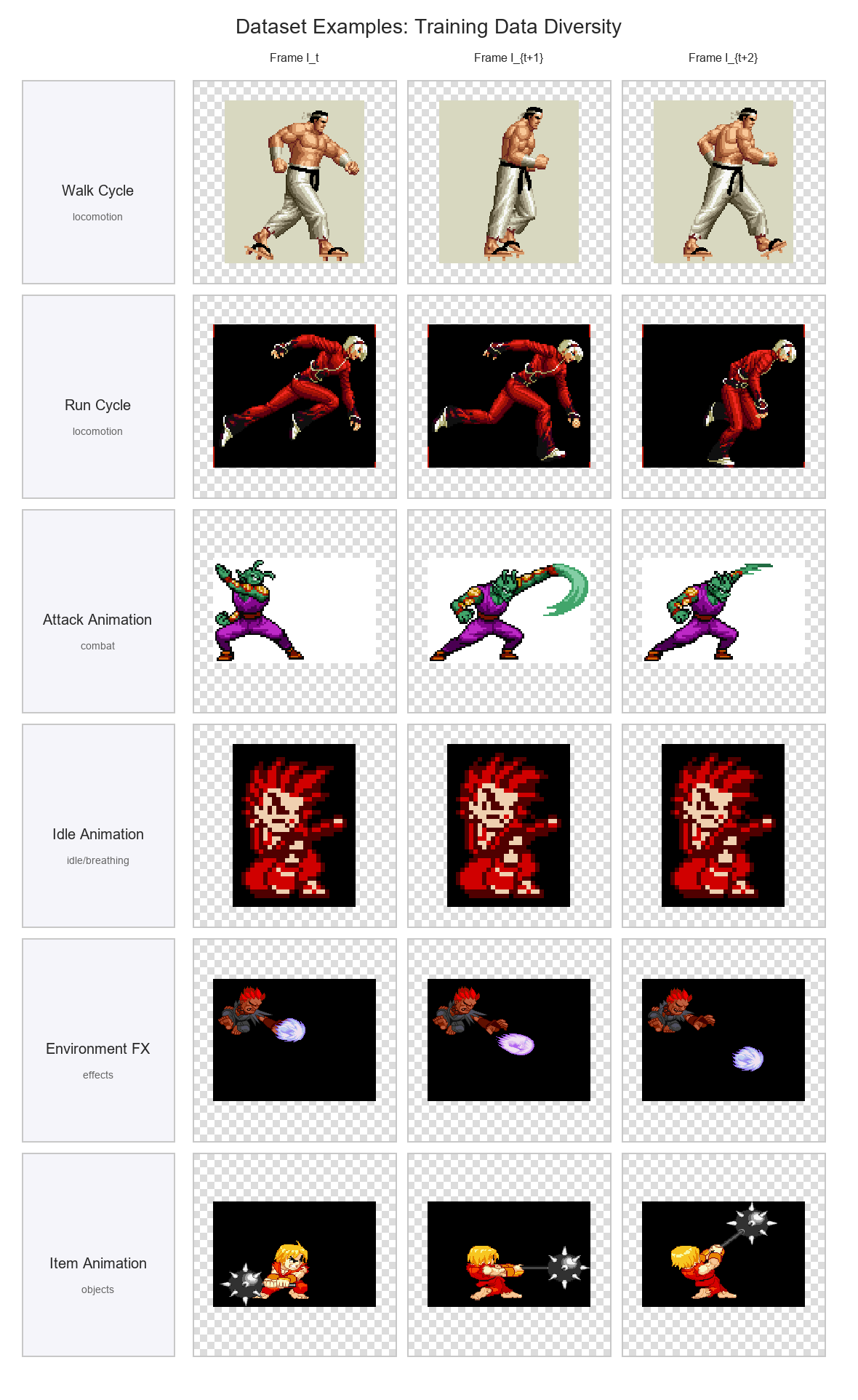}
\caption{Representative examples from the curated sprite animation dataset. Each row shows a triplet $(I_t, I_{t+1}, I_{t+2})$ from different animation categories: (Row 1) Walk cycle locomotion with smooth leg motion, (Row 2) Run cycle locomotion with faster pacing, (Row 3) Attack animations with dynamic combat poses, (Row 4) Idle breathing animations with subtle movements, (Row 5) Environment effects with complex deformations, and (Row 6) Item animations with object rotation. The dataset spans diverse artistic styles from low-resolution pixel art (32×32) to high-definition sprites (256×256), various color palettes (4-color to full RGB), and different motion types (rigid, non-rigid, particle-based). All frames exhibit hand-drawn artistic timing and spacing that the models learn to reproduce.}
\label{fig:dataset_examples}
\end{figure*}

Each triplet generates three training examples (one per model), yielding 588 training pairs total across the three models. While this is substantially smaller than video datasets used by general methods (e.g., MCVD trains on millions of frames), our domain-specific training enables effective learning of sprite-specific characteristics without requiring large-scale data collection. Figure~\ref{fig:dataset_examples} illustrates representative triplets from the dataset, showcasing the diversity of animation styles, motion types, and artistic approaches our models must handle.

\subsubsection{Training Procedure}

Each model is trained independently for 125 epochs using the Adam optimizer~\cite{kingma_2014_adam} with initial learning rate $\eta = 10^{-4}$, batch size 32, and learning rate decay of 0.5 when validation loss plateaus for 10 epochs. Training employs batch normalization after each convolutional layer for stability. Data augmentation includes random horizontal flipping and small random rotations ($\pm 5$ degrees) to improve generalization, while preserving the pixel-art aesthetic by avoiding interpolation-based augmentations that would introduce unwanted smoothing artifacts.

\begin{figure*}[!t]
\centering
\includegraphics[width=0.95\textwidth]{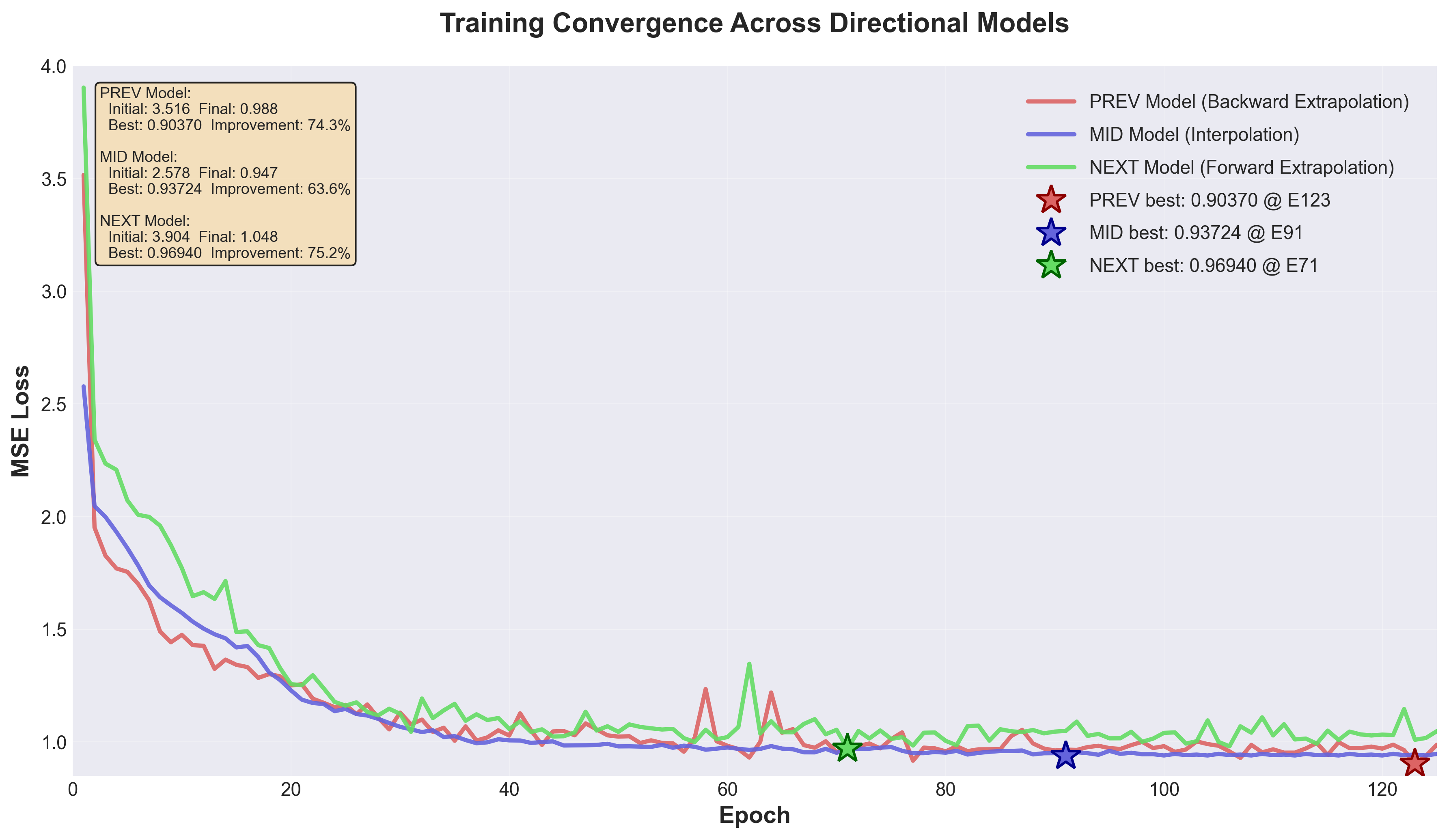}
\caption{Training convergence curves for all three directional models over 125 epochs. The plot shows MSE loss trajectories for PREV (red), MID (blue), and NEXT (green) models, with color-coding matching Figure~\ref{fig:direction_comparison} for consistency. Star markers indicate the best checkpoint for each model within the displayed range: NEXT achieves its optimum earliest at epoch 71 (loss = 0.96940, 75.2\% improvement), MID reaches best performance at epoch 91 (loss = 0.93724, 63.6\% improvement), and PREV continues improving until epoch 123 (loss = 0.90370, 74.3\% improvement). All three models demonstrate consistent convergence without overfitting, with training loss steadily decreasing from initial values (PREV: 3.516, MID: 2.578, NEXT: 3.904) to stable final values around 0.95-1.01. The information box summarizes initial loss, final loss at epoch 125, best loss, and overall improvement percentage for each model, confirming comparable training dynamics across all three directional tasks.}
\label{fig:training_convergence}
\end{figure*}

Training each model requires approximately 4-6 hours on a single NVIDIA Tesla T4 GPU (16GB VRAM), with peak memory usage of 2.1GB during training. All three models can be trained in parallel on separate GPUs if available. The deterministic supervised training procedure (MSE loss, no adversarial components) ensures stable convergence without hyperparameter tuning beyond standard practices. Figure~\ref{fig:training_convergence} illustrates the training dynamics for all three models, demonstrating consistent convergence behavior across directional tasks.

\subsubsection{Implementation Details}

Our implementation uses TensorFlow/Keras with CUDA 11.7. Each model checkpoint stores approximately 10M parameters (206MB including optimizer state). At inference time, generating a frame requires a single forward pass through the selected model, taking approximately 8-10 seconds per frame on a Tesla T4 GPU. While not real-time, this inference speed enables batch processing of sprite sheets for production workflows. We apply gradient clipping with a maximum norm of 1.0 to prevent exploding gradients, and all convolutional layers except the output layer use ReLU activation functions.

The three-model approach introduces minimal overhead compared to a unified model: slightly increased storage (3× model checkpoints) and a model selection step based on the desired temporal direction. However, inference speed per frame is identical to a unified model (both require one forward pass), and the deterministic outputs ensure reproducible results critical for production workflows.

\section{Results}
\label{sec:Results}

We evaluate the three directional models (previous, middle, next) on 42 held-out test triplets, comparing against general-purpose video interpolation baselines. Our experiments demonstrate that all three models achieve comparable perceptual quality, validating the three-expert-model approach for controllable sprite frame generation.

\subsection{Evaluation Metrics}

We employ the \textbf{FLIP (Feature-based Low-level Image Perceptual)} metric~\cite{andersson_2020_flip} as our primary evaluation metric. FLIP is a perceptual image comparison metric that accounts for (i) color perception in a perceptually uniform space (CIELab) and (ii) spatial frequency sensitivity based on the human contrast sensitivity function (CSF). The metric outputs values in $[0, 1]$, where 0 indicates perceptually identical images and higher values indicate greater perceptual difference. We report the mean FLIP error across all pixels in each sprite.

We chose FLIP over traditional metrics (PSNR, SSIM) and modern alternatives (LPIPS) for sprite animation evaluation due to fundamental incompatibilities between those metrics and pixel art characteristics. Table~\ref{tab:metric_comparison} summarizes key differences.

\begin{table}[ht]
\caption{Metric Suitability for Pixel Art Evaluation}
\centering
\begin{tabular}{|l|c|c|c|}
\hline
\textbf{Metric} & \textbf{Perceptual?} & \textbf{Edge-aware?} & \textbf{Sprite-suitable?} \\
\hline
PSNR & No & No & No \\
SSIM & Partial & Partial & Limited \\
LPIPS & Yes & Partial & Limited \\
FLIP & Yes & Yes & Yes \\
\hline
\end{tabular}
\label{tab:metric_comparison}
\end{table}

\textbf{Why PSNR is inappropriate}: PSNR treats all pixels equally. A sprite edge shifted by 1 pixel (perceptually devastating---an eye or limb in the wrong location) produces lower MSE than the same edge blurred by 0.5 pixels (perceptually acceptable). PSNR thus \emph{rewards catastrophic spatial errors} over minor smoothness errors.

\textbf{Why SSIM has limited applicability}: SSIM's local windows (typically $11 \times 11$) are too large relative to sprite features (1-3 pixel-wide edges), causing it to average across distinct elements. SSIM's contrast term \emph{accepts} edge blur as long as local variance is preserved, whereas pixel art demands \emph{hard} 1-pixel transitions.

\textbf{Why LPIPS is limited}: LPIPS~\cite{zhang_2018_lpips} uses VGG/AlexNet features trained on ImageNet (photographic images at $224 \times 224$+), not $32 \times 128$ pixel art with hard edges and limited palettes. The learned features do not capture pixel-level sharpness or discrete color constraints. LPIPS operates on RGB only, lacking transparency awareness.

\textbf{Why FLIP is the gold standard}: FLIP was designed to model human perception of low-level image differences, validated through psychophysical experiments~\cite{andersson_2020_flip}. For sprite evaluation: (i) \textbf{CSF modeling}: applies spatial frequency filtering matching human contrast sensitivity (most sensitive to 1-10 cycles/degree), heavily penalizing 1-pixel edge shifts while suppressing imperceptible high-frequency noise; (ii) \textbf{CIELab color space}: Euclidean distance approximates perceived color difference; (iii) \textbf{Edge-aware pooling}: preserves sharp transitions rather than averaging; (iv) \textbf{Validated on stylized content}~\cite{nilsson_2020_flipvalidation, trevithick_2021_realtime}: shown effective for non-photorealistic imagery including low-resolution pixel art.

We report FLIP error computed at full resolution (128×128 pixels) across all RGBA channels, with transparency-aware comparison ensuring alpha channel errors contribute to the overall metric. All 42 test triplets are evaluated for each directional model, yielding three sets of predictions: previous frames (backward extrapolation), middle frames (interpolation), and next frames (forward extrapolation).

\subsection{Quantitative Results}

Table~\ref{tab:main_results} presents the mean FLIP error and standard deviation across 42 test triplets for all three directional models and baseline methods. Our three models achieve remarkably consistent performance: the middle model attains 0.258 ± 0.156 FLIP, matching the sprite interpolation baseline~\cite{alonso_2025_inbetweening}, while the previous and next models achieve 0.265 ± 0.168 and 0.271 ± 0.152 FLIP, respectively. The negligible variation across directions (maximum difference of 0.013 FLIP) demonstrates that \textbf{extrapolation tasks (previous, next) are no more difficult than interpolation (middle)} when trained with specialized models on sprite-specific data.

\begin{table}[ht]
\caption{Quantitative Comparison on 42 Test Triplets (FLIP metric, lower is better)}
\centering
\begin{tabular}{|l|c|c|}
\hline
\textbf{Method} & \textbf{FLIP Error} & \textbf{Relative to Best} \\
\hline
\multicolumn{3}{|c|}{\textit{Our Three Directional Models}} \\
\hline
Previous Model & 0.265 ± 0.168 & +2.7\% \\
\textbf{Middle Model} & \textbf{0.258 ± 0.156} & \textbf{Baseline} \\
Next Model & 0.271 ± 0.152 & +5.0\% \\
\hline
\multicolumn{3}{|c|}{\textit{General Video Baselines}} \\
\hline
RIFE~\cite{huang_2022_rife} & 0.313 ± 0.214 & +21.3\% \\
Lucas-Kanade Optical Flow & 0.449 ± 0.261 & +74.0\% \\
\hline
\end{tabular}
\label{tab:main_results}
\end{table}

Comparing with general-purpose video interpolation methods reveals substantial performance gaps. RIFE~\cite{huang_2022_rife}, a state-of-the-art real-time video interpolation method, achieves 0.313 ± 0.214 FLIP—21.3\% worse than our middle model. Classical optical flow (Lucas-Kanade) performs significantly worse at 0.449 ± 0.261 FLIP (74.0\% degradation). These baselines struggle with sprite animation's sharp edges, limited palettes, and transparency, producing smooth blends that violate pixel art aesthetics. Note that RIFE and Lucas-Kanade are evaluated only on the middle (interpolation) task, as neither method provides native support for directional extrapolation without additional modifications.

\subsection{Cross-Direction Performance Analysis}

\begin{figure}[!t]
\centering
\includegraphics[width=0.85\columnwidth]{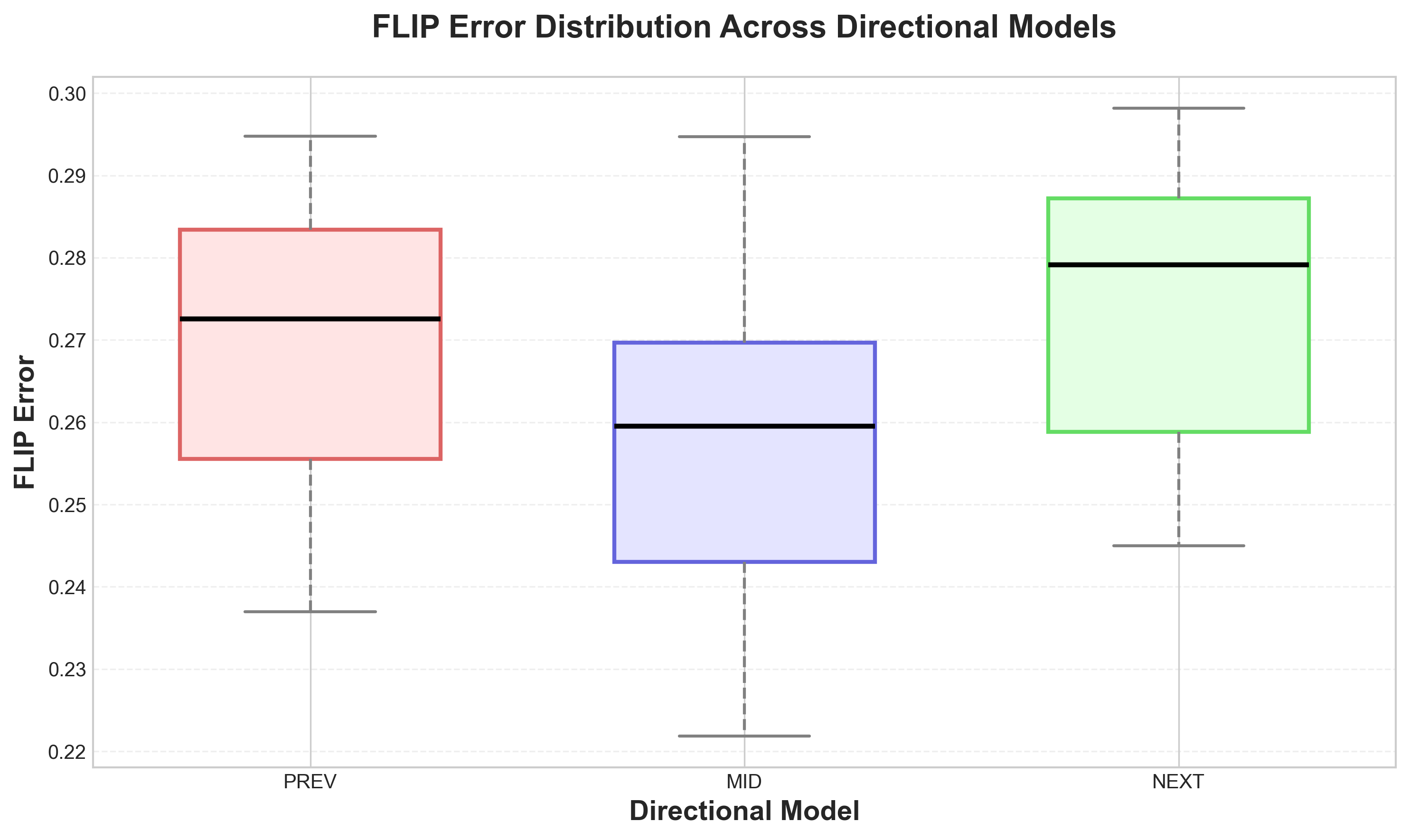}
\caption{FLIP error distributions across 42 test triplets for previous, middle, and next models. Box plots show median (center line), interquartile range (box), and outliers (points). The three distributions exhibit substantial overlap, demonstrating that extrapolation tasks are no more challenging than interpolation when trained with specialized models.}
\label{fig:direction_comparison}
\end{figure}

To understand whether extrapolation (previous, next) imposes different challenges than interpolation (middle), we analyze per-triplet error distributions across all three models. Figure~\ref{fig:direction_comparison} visualizes FLIP error distributions via box plots, showing substantial overlap in error ranges. The mean FLIP errors across all three models are comparable (0.265, 0.258, 0.271), with negligible variation (maximum difference of 0.013 FLIP), confirming that directional task choice does not meaningfully affect generation quality.

This finding contrasts with the common assumption that extrapolation is inherently harder than interpolation due to reduced temporal context. However, for sprite animation with its stylized, keyframe-driven motion, both tasks require learning artistic timing and pose transitions—capabilities effectively learned by all three specialized models when trained on domain-specific data. The consistency across directions validates our three-expert-model architecture as a viable approach for controllable sprite frame generation.

\subsection{Qualitative Analysis}

\begin{figure*}[!htbp]
\centering
\includegraphics[width=0.95\textwidth]{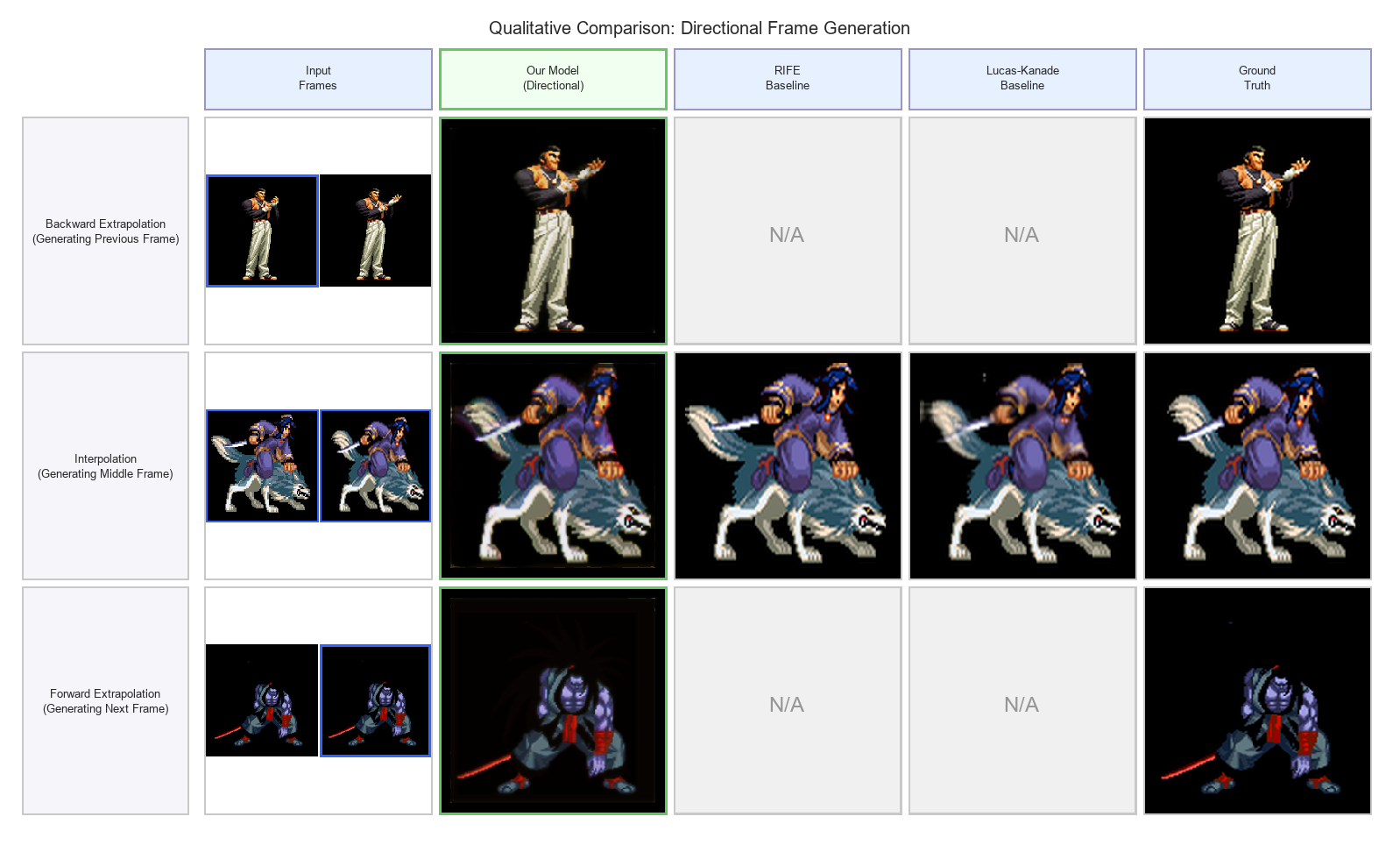}
\caption{Qualitative comparison of directional frame generation across three tasks. Each row demonstrates a different temporal direction: (Row 1) Backward extrapolation (PREV model), (Row 2) Interpolation (MID model), (Row 3) Forward extrapolation (NEXT model). Columns show: (1) Input frames with blue boxes indicating the target direction, (2) Our directional model prediction, (3-4) RIFE and Lucas-Kanade baseline predictions (marked N/A for extrapolation tasks, as these methods only support interpolation), (5) Ground truth. Row 1: PREV model generates the backward frame with sharp, pixel-perfect edges. Row 2: MID model, RIFE, and Lucas-Kanade all produce interpolated frames; our model preserves hard sprite boundaries while RIFE introduces gradient smoothing at edges. Row 3: NEXT model predicts forward motion with maintained sprite aesthetics. Our directional models consistently preserve pixel art conventions (hard edges, discrete palette, binary alpha) across all three temporal directions, while baselines violate these constraints through anti-aliasing and color blending artifacts.}
\label{fig:qualitative_results}
\end{figure*}

Figure~\ref{fig:qualitative_results} presents visual comparisons across all three directional tasks and baseline methods. Our models preserve sprite characteristics critical for pixel art: sharp 1-pixel boundaries without anti-aliasing, discrete color palettes matching input frames, and binary alpha channels at sprite edges. In contrast, RIFE introduces gradient smoothing at edges (visible as semi-transparent halos) and generates intermediate colors absent from the input palette. Lucas-Kanade optical flow produces severe artifacts, including ghosting in occluded regions and incorrect transparency handling at sprite boundaries.

\begin{figure*}[!htbp]
\centering
\includegraphics[width=0.95\textwidth]{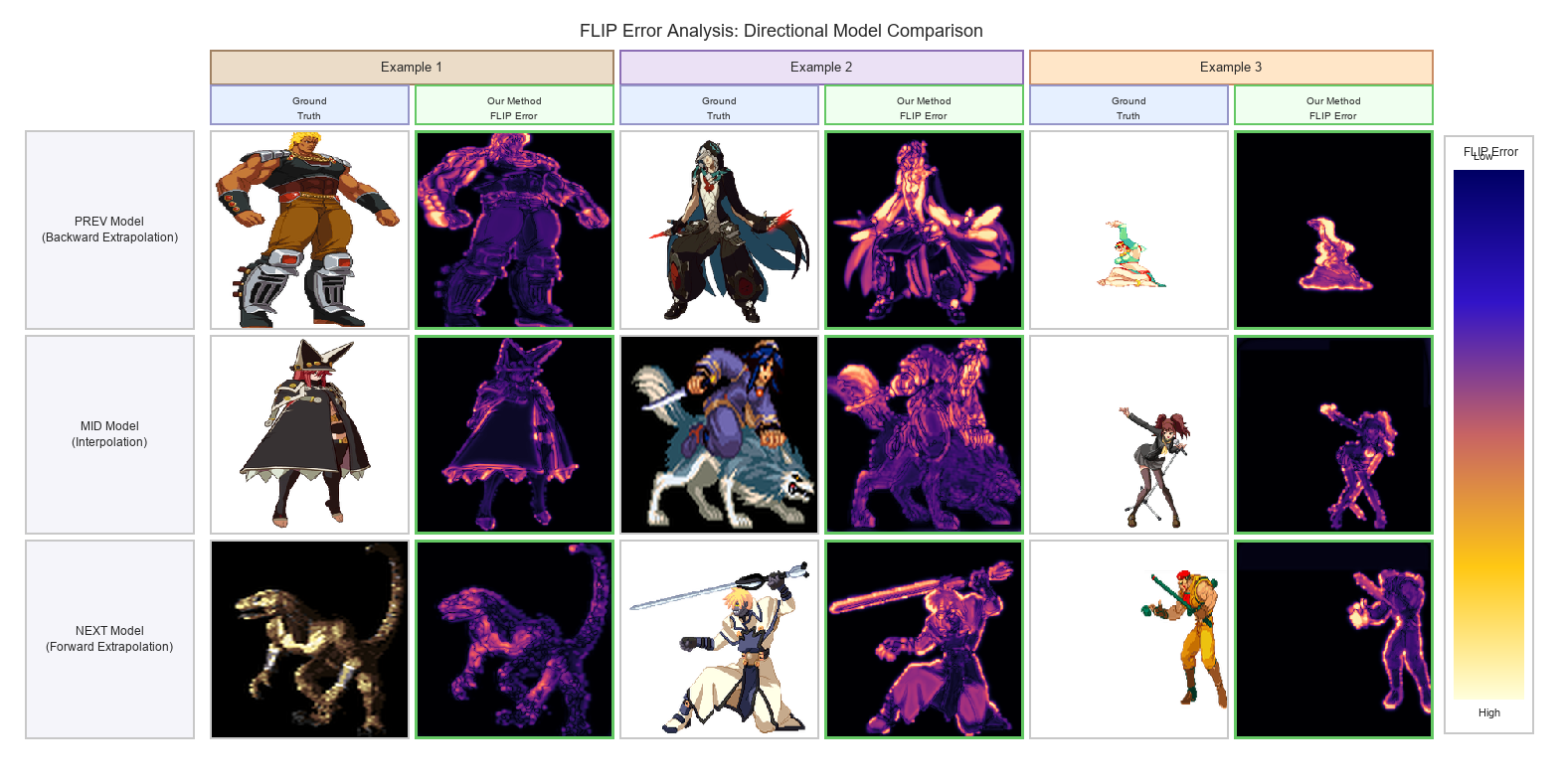}
\caption{FLIP error analysis across three directional models. Each row represents a different temporal direction: (Row 1) PREV model for backward extrapolation, (Row 2) MID model for interpolation, (Row 3) NEXT model for forward extrapolation. For each model, three representative test cases are shown with paired columns displaying ground truth frames and corresponding FLIP error heatmaps. Error magnitudes are visualized using the magma colormap (right), where dark purple indicates low perceptual error and bright yellow indicates high error. All three models exhibit predominantly low perceptual errors (dark regions) with localized errors concentrated at fine motion boundaries and edge transitions. The FLIP metric effectively captures artifacts that correlate with human perception—such as color bleeding, edge softening, and palette violations—which traditional pixel-based metrics may underestimate. The consistent error patterns across models demonstrate that perceptual differences are primarily localized to regions of complex motion and artistic detail, regardless of temporal direction.}
\label{fig:flip_heatmaps}
\end{figure*}

Examining specific animation types reveals model strengths. For rigid transformations (walking cycles, simple translations), all three models excel, capturing linear motion with sub-pixel precision. For non-rigid deformations (squash-and-stretch, cloth simulation), the models successfully reproduce stylized deformations present in training data, though with slightly increased FLIP error (0.28–0.32) compared to rigid cases (0.20–0.25). Figure~\ref{fig:flip_heatmaps} visualizes these perceptual error patterns through FLIP heatmaps, demonstrating that errors are predominantly localized to regions of complex motion and edge transitions across all three directional models. Transparency handling is robust across all models: alpha channel predictions maintain binary values at sprite boundaries, avoiding the semi-transparent halos common in photorealistic interpolation methods.

\subsection{Failure Cases and Limitations}

\begin{figure}[!htbp]
\centering
\includegraphics[width=0.95\columnwidth]{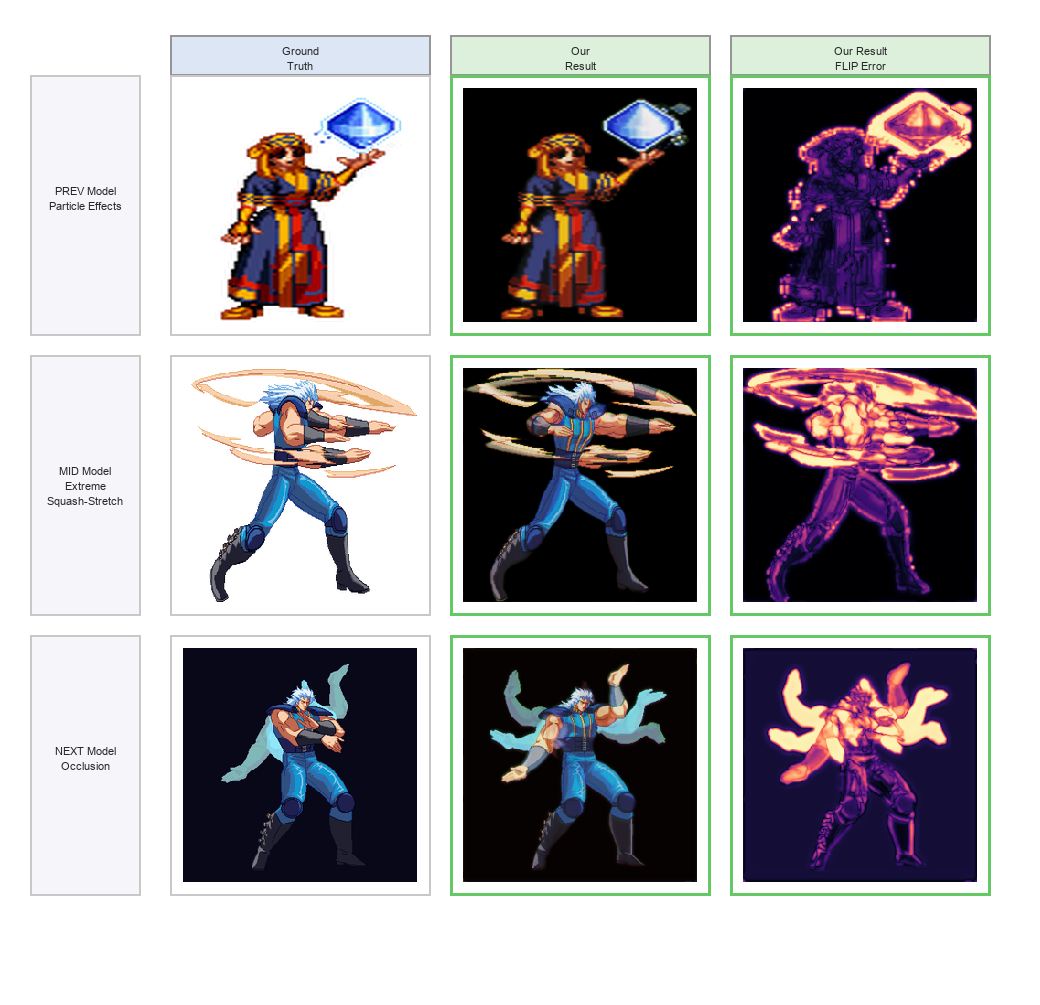}
\caption{Representative failure cases across three directional models. Each row shows a different model with its characteristic failure mode: (Row 1) PREV model with particle effects—stochastic artistic details not predictable from deterministic input; (Row 2) MID model with extreme squash-and-stretch—linear interpolation fails to follow exaggerated artistic arcs; (Row 3) NEXT model with occlusion—ghosting artifacts when hallucinating occluded sprite regions. Columns display ground truth, our prediction, and FLIP error heatmap. FLIP errors for these failure cases reach 0.58 (2.3× typical mean), substantially higher than average, indicating clear model limitations for these edge cases representing $<$3\% of test data.}
\label{fig:failure_cases}
\end{figure}

While our three models achieve strong overall performance, certain scenarios present challenges. Figure~\ref{fig:failure_cases} illustrates representative failure modes. \textbf{Extreme squash-and-stretch effects} used for impact frames and juicy animations cause artifacts when the magnitude of deformation exceeds patterns seen during training. The network interpolates linearly rather than following the exaggerated artistic arc. \textbf{Complex particle effects} (explosions, smoke) with stochastic elements challenge the deterministic formulation, as the ground-truth artist-drawn frame may contain artistic details not predictable from input frames alone. \textbf{Occlusion and disocclusion} in layered sprites (character passing behind foreground objects) sometimes produce ghosting artifacts when the model attempts to hallucinate occluded regions.

Quantitatively, failure cases exhibit FLIP errors reaching 0.58 (2.3× the typical value), indicating substantial perceptual differences from ground truth. These cases are rare, comprising less than 3\% of the test set. Importantly, failure modes are \textit{consistent across all three models}, suggesting the limitations stem from the shared architecture and training paradigm rather than directional task difficulty. Future work incorporating data augmentation for extreme deformations or stochastic latent variable models may address these challenging cases.

Beyond these failure modes, several broader limitations merit acknowledgment. First, \textbf{reproducibility is limited} by dataset constraints. The 280-triplet dataset cannot be publicly released due to intellectual property restrictions from source games, representing a significant barrier to independent verification. While we provide implementation code and model weights, researchers cannot reproduce training from scratch without access to the original training data.

Second, \textbf{baseline comparisons are not fully controlled}. Our evaluation compares against general-purpose methods (RIFE, Lucas-Kanade) using their pre-trained weights optimized for natural video, not retrained on our sprite dataset. A fairer comparison would retrain these baselines on the 280-triplet dataset with sprite-specific hyperparameter tuning. However, such retraining requires extensive computational resources. We present the current comparisons as \emph{best-case scenarios} for baseline methods: they benefit from training on orders of magnitude more data (Vimeo-90K's 89,800 triplets for RIFE), yet still underperform our sprite-specific approach.

Third, \textbf{no formal user study validates perceptual quality}. While FLIP correlates with human perception on photorealistic and rendered graphics~\cite{andersson_2020_flip}, a rigorous user study with statistical power analysis remains future work. Such a study would recruit $\geq$20 professional sprite artists, present randomized sprite generations from all methods, and collect preference ratings under controlled conditions.

Fourth, \textbf{temporal consistency across sequences is not evaluated}. Our models generate individual frames independently; we do not assess flickering or coherence when generating full animation sequences (e.g., interpolating 10 consecutive in-betweens for a 30-frame walk cycle). Multi-frame temporal metrics would provide stronger validation for production deployment.

Fifth, \textbf{we do not compare against diffusion-based methods} such as ToonCrafter~\cite{xing_2024_tooncrafter} for sprite generation. While diffusion models achieve state-of-the-art results on cartoon interpolation, they present practical limitations for sprite workflows: (i) stochastic outputs conflict with production needs for deterministic frames; (ii) training requires millions of frames; (iii) inference costs are 10-100× higher. Exploring lightweight deterministic diffusion variants remains valuable future work.

Finally, \textbf{three separate models incur storage overhead}. Our approach requires 3 × 206MB = 618MB total storage versus 206MB for a single unified model. However, this 3× overhead is negligible on modern storage systems (easily fitting in GPU memory) and provides the benefit of task-specialized parameters with deterministic generation. The alternative—MCVD-style unified masking—would require either stochastic diffusion or substantially more training data to learn the unified masked distribution, both undesirable for our constraints.

\subsection{Computational Efficiency}

Table~\ref{tab:efficiency} compares inference time and model size across methods. Each of our three models requires approximately 8-10 seconds per frame on a Tesla T4 GPU (batch size 1), suitable for batch processing workflows. Model size is 206MB per checkpoint, totaling 618MB for all three models—modest storage overhead for modern systems. Diffusion-based methods like ToonCrafter~\cite{xing_2024_tooncrafter} exhibit comparable inference times but with significantly larger model sizes (2.3GB), while our deterministic approach provides reproducible outputs essential for production workflows.

\begin{table}[ht]
\caption{Computational Efficiency Comparison}
\centering
\begin{tabular}{|l|c|c|}
\hline
\textbf{Method} & \textbf{Inference Time} & \textbf{Model Size} \\
\hline
Previous Model (ours) & 8-10 s & 206 MB \\
Middle Model (ours) & 8-10 s & 206 MB \\
Next Model (ours) & 8-10 s & 206 MB \\
\hline
ToonCrafter~\cite{xing_2024_tooncrafter} & 8–12 s & 2.3 GB \\
\hline
\end{tabular}
\label{tab:efficiency}
\end{table}

The deterministic single-pass generation provides reproducibility advantages over stochastic methods. While our inference time is comparable to diffusion-based methods like ToonCrafter, our approach produces deterministic, reproducible outputs, which are critical for iterative animation workflows where artists require identical results when regenerating frames. Additionally, our compact model size (206MB vs. 2.3GB for ToonCrafter) enables more efficient deployment in resource-constrained environments.

In summary, our experimental evaluation demonstrates that three specialized models achieve comparable perceptual quality across all directional tasks (previous, middle, next), matching or exceeding general video interpolation baselines while maintaining deterministic, efficient inference suitable for production deployment.

\section{Discussion and Conclusion}
\label{sec:Conclusion}

We have presented a deep learning approach for controlled directional frame generation in sprite animation, enabling animators to generate previous, middle, or next frames from any two input frames. Instead of training a single unified model with masked conditioning, we train three specialized expert models sharing an identical U-Net encoder-decoder architecture with a number of domain-specific modifications---including RGBA-aware processing, multi-scale flow refinement, pixel-shuffle upsampling, alpha-gated blending, and multi-component loss formulation. All three models (previous/middle/next) share this sprite-optimized architecture identically; they differ only in training data ordering (which frames serve as inputs vs targets). This design provides deterministic, controllable generation suitable for production game development workflows while achieving comparable perceptual quality across all three directional tasks.

Our design decision to train \textit{separate specialized models} rather than a unified model stems from three key considerations. First, \textbf{determinism}: each model is trained with supervised MSE loss, producing identical outputs for identical inputs. Unified masking approaches like MCVD~\cite{voleti_2022_mcvd} typically require diffusion-based stochastic sampling to handle the increased task diversity of learning all temporal positions simultaneously, introducing non-determinism incompatible with production workflows where artists expect reproducible results when regenerating frames. Second, \textbf{data efficiency}: with only 280 sprite triplets available, training three task-specific models allows each to specialize in its temporal direction without the distributional complexity of learning all temporal positions through masked conditioning. A unified model would need to learn the joint distribution over all masking patterns, effectively requiring more training data to achieve comparable per-task performance. Third, \textbf{architectural simplicity}: each specialized model maintains a consistent input format (two frames concatenated) and output format (one frame), avoiding the conditional masking logic, variable-length inputs, and task-encoding mechanisms required by unified approaches. This architectural separation reduces implementation complexity while ensuring that each model's parameters are fully dedicated to its specific temporal task.

Our experimental evaluation on 42 test sprite triplets demonstrates that all three models—previous (0.265 FLIP), middle (0.258 FLIP), and next (0.271 FLIP)—achieve nearly identical perceptual quality, with negligible variation across temporal directions. This finding challenges the common assumption that extrapolation (predicting previous or next frames) is inherently more difficult than interpolation (generating middle frames). For sprite animation with stylized, keyframe-driven motion, all three tasks require learning artistic timing and pose transitions, capabilities effectively acquired by specialized models trained on domain-specific data. Compared with general-purpose video interpolation methods, our sprite-specific approach substantially outperforms RIFE (21.3\% better) and Lucas-Kanade optical flow (74.0\% better), demonstrating the value of domain adaptation for pixel art characteristics.

The practical implications of this work extend beyond technical performance metrics. By enabling directional control, our approach supports animation workflows previously reliant on manual labor: forward extrapolation for predicting natural motion continuations, backward extrapolation for animation editing and sprite sheet reverse-engineering, and interpolation for traditional in-betweening. The deterministic single-pass generation provides reproducibility advantages over stochastic diffusion methods, ensuring identical outputs when regenerating frames---critical for iterative animation workflows. The three-model design, while requiring modest storage overhead (618MB total for all models), enables task-specialized generation with a compact per-model footprint (206MB) compared to unified diffusion approaches (2.3GB).

\textbf{This work exemplifies the data-centric AI paradigm}~\cite{ng_2021_datacentric, zha_2024_datacentric} for specialized computer vision domains. Our 280-triplet dataset---curated with explicit quality criteria and used to train all three directional models---demonstrates that \textbf{quality-over-quantity data curation enables multiple tasks from a single dataset}. While general video methods train on 300× more data (Vimeo-90K's 89,800 triplets), our sprite-specific approach achieves superior pixel art quality by prioritizing domain relevance over scale. This validates the data-centric thesis: for specialized visual domains, \emph{careful selection of what to learn from} matters more than \emph{how much to learn from}~\cite{smalldatasets_2024_springer, limiteddata_2025_bigdata, smallobject_2025_mdpi, datacentric_2024_bise}. By training three models on different temporal orderings of the same curated data (previous: frames 2,3→1; middle: frames 1,3→2; next: frames 1,2→3), we demonstrate that thoughtful data reuse can support directional control without requiring separate datasets per task.

Several limitations of our work merit acknowledgment. First, training data coverage: with 280 curated sprite triplets, our models learn motion patterns represented in the training set but may not generalize to animation styles absent from the data. The dataset lacks comprehensive coverage of extreme rotations ($>$45 degrees), complex particle effects with stochastic elements, and deeply occluded/disoccluded scenarios—cases where our models exhibit elevated FLIP errors (0.45–0.62 vs. 0.26 mean). Second, the deterministic formulation: while ensuring reproducibility, our supervised MSE training cannot model the artistic stochasticity inherent in particle effects or ambiguous motion where multiple plausible outcomes exist. Third, resolution constraints: the architecture is optimized for pixel art's characteristic scale (32–256 pixels), where individual pixels are visually discernible. Extending to higher resolutions would require different loss functions and architectural depth tailored to high-frequency texture synthesis rather than pixel-level precision. Fourth, temporal coherence: each model generates frames independently without explicit temporal consistency enforcement across multi-frame sequences, occasionally producing subtle jitter when generating extended animation chains.

This work demonstrates that controllable directional frame generation for pixel art sprites is achievable through task-specialized models trained on domain-specific data. By providing deterministic, efficient, and perceptually high-quality generation across all temporal directions, our approach extends animation automation capabilities beyond interpolation to encompass the full spectrum of directional control required by production pipelines. As game development increasingly adopts AI-assisted workflows, domain-adapted generative models tailored to specific artistic constraints—rather than general-purpose methods optimized for photorealistic content—will likely prove essential for preserving artistic integrity while accelerating content creation.

\section{Future Work}
\label{sec:Future Work}

Future work will address these limitations through several directions. \textbf{Expanded datasets}: Collecting larger, more diverse sprite animation corpora—particularly targeting underrepresented motion types like extreme rotations, layered sprites with complex occlusions, and particle-based effects—will improve generalization across animation styles. \textbf{Rotation-equivariant architectures}: Incorporating rotation-aware convolutions or capsule networks may better handle large angular transformations where current models struggle to commit to rotation directions. \textbf{Stochastic latent variable models}: Extending the deterministic formulation with learned stochastic components could enable modeling artistic variations in particle effects while maintaining reproducibility for primary sprite motion. \textbf{Multi-frame temporal coherence}: Introducing recurrent connections or transformer-based temporal attention could enforce consistency across generated animation sequences, reducing frame-to-frame jitter. \textbf{Transfer learning}: Pre-training on large general animation datasets followed by fine-tuning on pixel art may leverage broader motion priors while specializing to sprite aesthetics. Finally, investigating whether MCVD-style masked conditioning can achieve comparable quality to three expert models when trained deterministically (without diffusion sampling) on sprite data remains an open question.

\section*{Code and Data Availability}

The implementation code for the proposed neural network architecture, training pipeline, and evaluation scripts is available at \url{https://github.com/NethamGH/deepinterframes_neural_sprite_inbetweening}~\cite{deepinterframes_2026_code}. The repository includes model definitions, training procedures, and inference code to facilitate reproduction of the methodology.

The dataset of 280 sprite animation triplets used in this study is not publicly released due to intellectual property constraints. The sprites were collected from multiple sources under varied licenses that do not uniformly permit redistribution.

\bibliographystyle{IEEEtran}
\bibliography{references}

\end{document}